\documentclass[useAMS,usenatbib]{mn2e}
\usepackage{graphicx}
\usepackage{url}

\title[2MPZ and LIGO]{Using the 2-MASS Photometric Redshift Survey to optimise LIGO Follow-Up Observations}
\author[Antolini \& Heyl]{Elisa Antolini$^{1}$, Jeremy S. Heyl$\thanks{Email:
    heyl@phas.ubc.ca; Canada Research Chair}^{2}$ \\
  $^{1}$Dipartimento di Fisica e Geologia, Universit\`a degli Studi di Perugia, I-06123 Perugia, Italia \\
  $^{2}$Department of Physics and Astronomy, University of British
  Columbia, 6224 Agricultural Road, Vancouver, BC V6T 1Z1, Canada\\
}
\begin{document}
\date{Accepted 2016 July 13. Received 2016 July 12; in original form 2016 February 24}

\pagerange{\pageref{firstpage}--\pageref{lastpage}} \pubyear{2016}

\maketitle

\label{firstpage}

\begin{abstract}
  The initial discovery of LIGO on 14 September 2015 was the inspiral
  merger and ring-down of the black hole binary at a distance of about
  500~Mpc or a redshift of about 0.1.  The search for electromagnetic
  counterparts for the inspiral of binary black holes is impeded by
  coarse initial source localisations and a lack of a compelling model
  for the counterpart; therefore, rapid electromagnetic follow-up is
  required to understand the astrophysical context of these sources.
  Because astrophysical sources of gravitational radiation are likely
  to reside in galaxies, it would make sense to search first in
  regions where the LIGO-Virgo probability is large and where the
  density of galaxies is large as well.  Under the 
  assumption that the probability of a gravitational-wave event from a
  given region of space is proportional to the density of galaxies
  within the probed volume, one can calculate an improved localisation
  of the position of the source simply by multiplying the LIGO-Virgo
  skymap by the density of galaxies in the range of redshifts.  We
  propose using the 2-MASS Photometric Redshift Galaxy Catalogue for
  this purpose and demonstrate that using it can dramatically reduce
  the search region for electromagnetic counterparts.
\end{abstract}
\begin{keywords}
  gravitational waves: Physical Data and Processes --
  galaxies: distances and redshifts: Galaxies --
  methods: observational: Astronomical instrumentation, methods, and techniques
\end{keywords}
\section{Introduction}

LIGO has recently begun to detect gravitational wave events from the
local Universe \citep{PhysRevLett.116.061102}.  During these initial
years of gravitational astronomy, the localisation of the candidate
events on the sky is coarse with the ninety-percent confidence regions
covering hundreds or even thousands of square degrees .
\citep{2014ApJ...789L...5K,2014ApJ...795..105S,2015ApJ...804..114B,2016LRR....19....1A}.
Finding an electromagnetic counterpart to these candidate
gravitational-wave events will be crucial to understand the host
environment, the evolution of the progenitor and to provide tests of
cosmology by yielding measurement of the redshift of the source.  The
ideas of how the the electromagnetic counterparts would appear are
varied and uncertain. There has been substantial consideration of the
electromagnetic transients associated with the mergers of binaries
that include a neutron star
\citep[e.g.][]{2016PhRvD..93b4011E,2016arXiv160107711K,2016arXiv160100017D,
  2015arXiv151205435F,2015ApJ...814L..20M,2015PhRvD..92d4028K,
  2015arXiv150807939S,2015arXiv150807911S} However,the first
discovered gravitational wave event (GW150914) was almost certainly
the merger of binary black holes, so the appearance and duration of
the electromagnetic counterparts are especially uncertain with only a
few models
\citep[e.g.][]{2015PhRvL.115n1102G,2015MNRAS.452.3419M,2016MNRAS.457..939C,2016ApJ...817..183Y}.
Consequently, rapid electromagnetic follow-up of a large portion of
the probable region would increase the chance of success in finding a
counterpart.  Over the large search regions and over the span of days
or weeks, many electromagnetic transients typically occur, and with
the wide variety of models it will be difficult to associate
unambiguously a particular electromagnetic event with a candidate
gravitational-wave event.

The purpose of this paper is to present a strategy to alleviate both
of these issues; that is, to reduce both the search region and the
time required to plan and begin observations.  We follow the approach of
\citet{2015arXiv150803608G} to develop a galaxy catalogue to guide the
observational plan \citep[see
  also][]{PhysRevD.82.102002,0004-637X-784-1-8,Ghosh2015,Fan2015,2015ApJ...801L...1B}.
However, our goal here is to develop a nearly complete catalogue at
the expense of having less accurate estimates of the redshifts of the
galaxies within the catalogue.  The accuracy of the galaxy distances
needs to be only as good as the distance estimates of the
gravitational-wave events.  Additionally we will outline a
straightforward and rapid technique to generate a nearly optimal
observing plan to follow up the events rapidly (i.e. within a few
seconds of the trigger).

\section{Bayesian Approach to Follow-Up}

Because we will be interested in the rapid follow-up of candidate
gravitational-wave events, we will be focused on the rapid Bayesian
reconstruction outlined by \citet{2015arXiv150803634S}, BAYESTAR. 
BAYESTAR yields a probability map on the sky in
the form of a HEALPix map \citep{2005ApJ...622..759G} where each pixel
contains the posterior probability $P(m|d)$ of a particular model
parameterised by the position on the sky conditioned on the observed
data (i.e. the observed strains on the LIGO and Virgo
interferometers).  To plan an observing strategy one would like the
probability of a particular model (i.e. position on the sky).
We have from Bayes's theorem
\begin{equation}
  P(\mathrm{position}|\mathrm{data}) = \frac{P(\mathrm{position})
    P(\mathrm{data}|\mathrm{position})}{P(\mathrm{data})}.
  \label{eq:1}
\end{equation}
If we make the additional mild assumption that gravitational-waves
originate from nearby galaxies, the probability of a given position on
the sky naturally is proportional to the density (or perhaps the
luminosity density) of galaxies in that direction integrated over
distance range determined from the modelling of the gravitational
waveform.  Of course, these distance estimates will usually have large
uncertainties so the distance range over which to integrate the
galaxy density distribution will also be large, so highly accurate
redshift information is not needed to construct
$P(\mathrm{position})$.

Furthermore, because we will ultimately be interested in which fields
to observe (not which particular galaxies), accurate positions are not
required in the construction of $P(\mathrm{position})$. Because the
LIGO probability maps are sampled on a HEALPix grid, it is convenient
to sample $P(\mathrm{position})$ also as a HEALPix grid with each
pixel covering about the same solid angle as the field of view of the
telescope of interest or the BAYESTAR map.  We choose a HEALPix map
with $\mathtt{NSIDE}=512$ or about 50 square arcminutes per pixel, so
positions no more accurate than arcminutes are required.  The key to
generate the observing plan rapidly is to calculate the required
galaxy density maps beforehand in principle at the desired resolution
(this optimisation only speeds the process up slightly) for the
distance ranges of interest.  With the arrival of an alert, all that
is required is to calculate Eq.~(\ref{eq:1}) using the HEALPix maps,
resample to the scale of the telescope, renormalise the probability,
sort the pixels from most likely to least and output the positions to
cover a given amount of cumulative probability (this entire process
takes typically less than one second).

\section{Galaxy Catalogues}

To gain a picture of the local Universe, our focus will be the
completeness of the data rather than the accuracy of the distances and
positions.  The Census of the Local Universe
\citep[CLU;][]{2015arXiv150803608G} combines several redshift surveys
(2dF Galaxy Redshift Survey \citealt{2002MNRAS.336..907N}; the
Millenium Galaxy Catalog,
\citealt{2003MNRAS.344..307L,2005MNRAS.360...81D}; and the 2MASS
Reshift Survey \citealt{2012ApJS..199...26H}) that cover a large
portion of the sky, but at various depths and attempt to increasing
the completeness of the sample by using only the galaxies near the
upper-end of the luminosity function ({\em i.e.}  $L\sim L_*$).  This
may optimise the strategy to discover neutron-star binaries whose
abundance is probably proportional to the total number of stars in a
galaxy, and galaxies near $L_*$ dominate the stellar mass in the
Universe \citep[however see][for caveats to this
  assumption]{2010ApJ...725.1202L}.  On the other hand, the discovery
that binary black holes with large masses dominate the initial
detections indicates that focusing the search on massive galaxies
might not be the best strategy to discover the electromagnetic
counterparts to the first sources.  After all such large black holes
have not been found so far in our approximately $L_*$-galaxy, the
Milky Way, or our neighbour, Andromeda.  In fact theoretical arguments
indicate that the production of such massive black holes results from
the evolution of massive stars in low metallicity galaxies
\citep{2016ApJ...818L..22A,2016arXiv160203790E} which are typically
small in the local Universe \citep[e.g.][]{1997MNRAS.285..613H}. Our
goal is to have a nearly complete survey that attempts to be unbiased
with respect to the mass of the galaxy.

We follow in spirit the work of \citet{2004PASA...21..396J} who used
The Two Micron All Sky Survey extended source catalogue \citep[2MASS
  XSC,][]{2000AJ....119.2498J,2006AJ....131.1163S}, and the assumption
that all galaxies have the same $K_s-$band luminosity of around $L_*$
to estimate distances to each galaxy and create sky maps of the local
Universe.  A substantial fraction of 2MASS has measured redshifts
\citep[e.g][]{2012ApJS..199...26H}.  \citet{2014ApJS..210....9B}
combined the photometric data from 2MASS XSC with additional
photometry the mid-infrared from WISE \citep{2010AJ....140.1868W} and
the optical from SuperCOSMOS
\citep{2001MNRAS.326.1315H,2001MNRAS.326.1295H,2001MNRAS.326.1279H}.
Using this multiband photometry, they trained neural networks using
measured spectroscopic redshifts from SDSS
\citep{2012ApJS..203...21A,2014ApJS..211...17A}, 2dF
\citep{2001MNRAS.328.1039C,2003astro.ph..6581C}, 6dF
\citep{2004MNRAS.355..747J,2009MNRAS.399..683J} and other catalogues
to determine photometric redshifts.  They also extend the photometric
redshift catalogue beyond the 2MASS XSC building a three-dimensional
map of the sky out to a redshift of nearly 0.2 well into the realm of
the first gravitational wave event.  The 2MASS Photometric Redshift
(2MPZ) catalogue contains over one million galaxies with a median
redshift of 0.1 with a typical scatter between the
photometric and spectroscopic redshifts (where both are known) of
$\sigma_z = 0.015$.

\begin{figure}
  \includegraphics[width=\columnwidth]{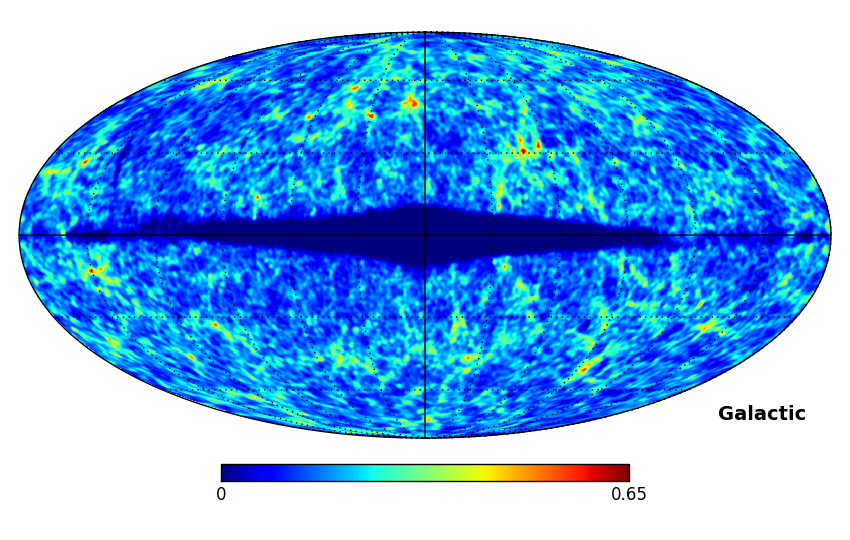}
  \caption{The relative surface density of galaxies in the 2-MASS Photometric Redshift Survey with photometric redshifts between 0.01 and 0.1, smoothed with a Gaussian of 0.6 degrees (0.01 radian).}
  \label{fig:galmap}
\end{figure}

Except for the most local binary-black-hole mergers and
neutron-star-black-hole mergers, the estimated distances from the
gravitational wave data itself have comparable errors to this, so this
catalogue is sufficiently accurate to calculate the surface density of
galaxies with the expected redshift range of a particular
gravitational-wave detection.  On the other hand, the
binary-neutron-star sources will have distances comparable to these
uncertainties.  Fortunately, for these nearby sources, there are
nearly uniform all-sky redshift surveys which would be more
appropriate for this task
\citep[e.g.][]{2000MNRAS.317...55S,2012ApJS..199...26H}.  Of course,
all of the techniques outlined here can be applied to these more
nearby catalogues to produce sky maps of even more nearby galaxies.
Here we will focus on galaxies with photometric redshifts between 0.01
and 0.1 from the 2MPZ catalogue as depicted in Fig.~\ref{fig:galmap}.
For closer galaxies the redshift error is significant (the mean error
is about 0.015), and the outer end of the range is both the median
redshift of the catalogue and the typical distances of the
binary-black-hole sources.  Although at the inner edge of the
catalogue the redshift error exceeds the redshift estimate, this does
not impede our goal of creating a catalogue of nearby galaxies to
look for counterparts to binary-black-hole mergers.

To produce this map we divided the sky into 3,145,728 regions (each of
about 45 square arcminutes, four ACS fields) using a HEALPix
\citep{2005ApJ...622..759G} tessellation with $\mathtt{NSIDE}=512$.
Each cell of the map simply contains the number of galaxies in the
2MPZ catalogue within the range of photometric redshift that lie
within that portion of sky.  We have consequently smoothed the map
with a Gaussian of 0.6 degrees or about 2~Mpc at a redshift of 0.05.
This smoothing accounts for the possibility that either the
binary-black hole has been kicked out of one of the catalogue galaxies
(1000~km/s for 1~Gyr yields 1~Mpc) or that the catalogue galaxies are
accommpanied by smaller galaxies that are absent from the catalogue
but cluster around those in the catalogue within a typical group scale
of 1~Mpc.  In prinicple we could smooth the maps more finely with
increasing distance; these results would necessarily be closer to the
unsmoothed map.

Typically no HEALPix pixel contains more than one galaxy
from the catalogue.  After smoothing we notice the large-scale
structure even when we have averaged over distance.  This demonstrates
the potential optimisation in the observing strategy by observing
fields with nearby galaxies. Depending on whether one believes that
the sources are associated with the visible portion of galaxies or may
travel some distance from the galaxy itself before the
gravitational-wave event, one would use either the raw galaxy counts
or the smoothed map.

Furthermore, our choice of weighting the fields simply by the number
of galaxies within each field is perhaps the most simple one.  Given
the type of event, one could use a map that gives small,
low-metallicity galaxies more weight or weigh the galaxies by their
mass or luminosity.  Of course, all of these possibilities would be
informed by one's prior knowledge of the source guided by
theoretical models and the hints from the waveform itself and give a
better estimate of $P(\mathrm{position})$.  The key is to calculate these
maps beforehand. 

\begin{figure}
  \includegraphics[width=\columnwidth,clip,trim=0 1in 0 0]{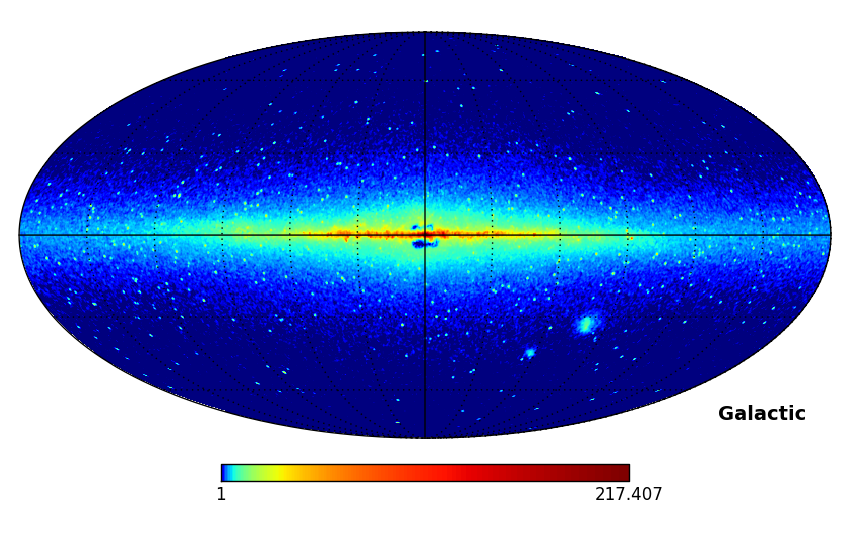}
  \caption{The relative surface brightness of stars in the 2-MASS
    Photometric Catalogue \citep{2006AJ....131.1163S} smoothed with a Gaussian
    of 0.03 degrees (0.005 radian).}
  \label{fig:starmap}
\end{figure}
There is a further structure apparent in the map, and this is the zone
of avoidance imposed by the disk and bulge of our Galaxy. Depending on
the nature of the follow-up ({\em e.g.} gamma-ray and radio)
observations it may make sense to include regions along the Galactic
plane where one thinks nearby galaxies should be. One can attempt to
probe the zone of avoidance \citep[e.g][]{2000AJ....120..298J} and
future 21-cm surveys like CHIME \citep{2014era..conf10102V} will also
probe the large-scale structure beyond the Galactic plane.  However,
the existing galaxy density map as depicted in Fig.~\ref{fig:galmap}
can yield an estimate of the structure obscured by the Galaxy.  The
technique that we will use in similar to that used by
\citet{2008StMet...5..289A} to inpaint the CMB anisotropies across the
Galactic plane.

\begin{figure}
  \includegraphics[width=\columnwidth,clip,trim=0 1in 0 0]{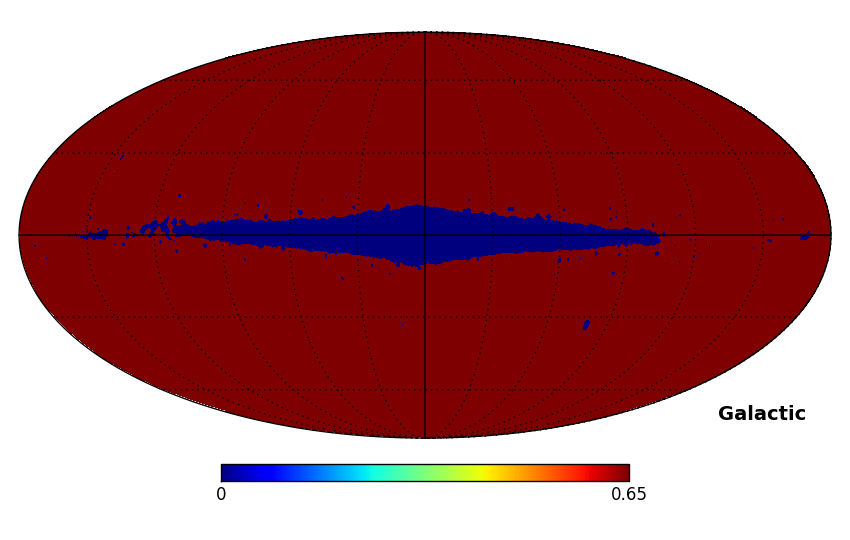}
  \includegraphics[width=\columnwidth]{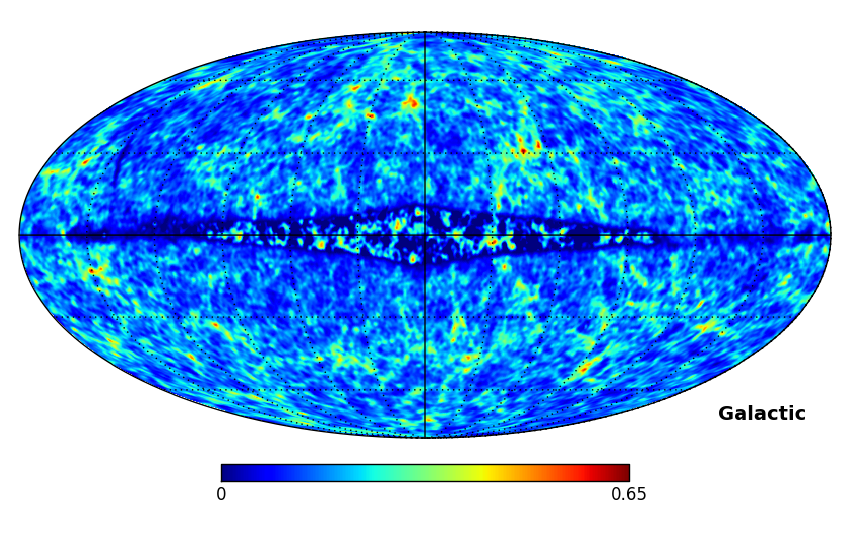}
  \includegraphics[width=\columnwidth]{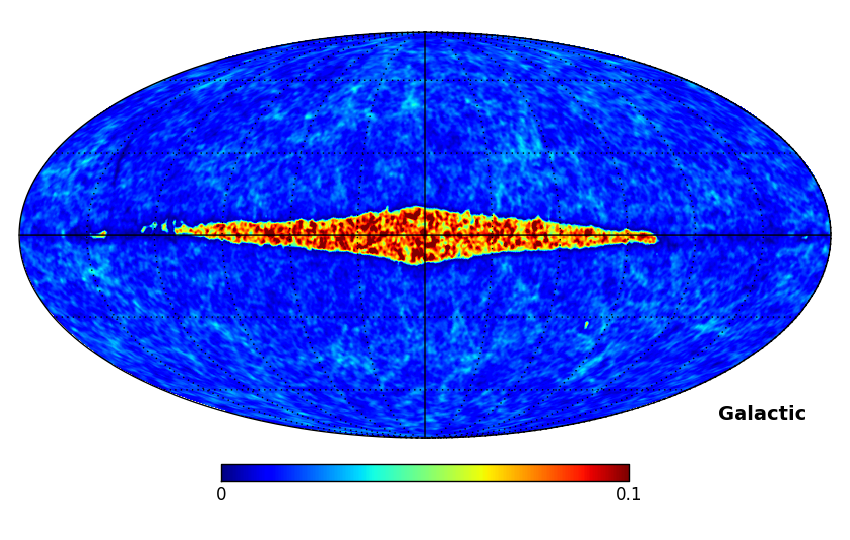}
  \caption{Upper: the mask used for the infilling procedure obtained
    by determining the regions where the galaxy density is less than
    one tenth of the mean or the star density lies above a given
    threshold (see text for details).  Middle: the infilled galaxy
    distribution. Lower: The standard deviation of the infilled map
    obtained by bootstrapping the galaxy catalogue.}
  \label{fig:infilling}
\end{figure}

Here we will determine the region masked by the Galaxy by finding the
region in which the density of galaxies is either less than one tenth
of the mean (from Fig.~\ref{fig:galmap}) or in which the density of
stars (from Fig.~\ref{fig:starmap}) is greater than a threshold that
accounts for the masking of the background galaxies due to the Large
Magellanic Cloud, a feature that is apparent in both figures.  Both of
these masks are nearly the same, so we combine them as depicted in the
upper panel of Fig.~\ref{fig:infilling}.  This region is much narrower
than the infilled region of the CMB in \citet{2008StMet...5..289A},
and furthermore the observed structures the galaxy map are typically
longer than the width of the mask, so we can reliably estimate the
hidden structures.  In spite of these differences the basic strategy
is similar.  We assume that the underlying galaxy map (behind the
Galaxy) is isotropic; therefore, it is natural to represent it as a
sum of spherical harmonics; furthermore, we can argue that a small
fraction of the components contain most of the power, {\em i.e.} the
representation of the underlying map is sparse, so we can use the
adaptive thresholding strategy of \citet{2007ITIP...16.2675B} to
estimate the underlying galaxy distribution.  \citet{2016AntolInfill}
give the details of the procedure as well as a variety of tests.

\begin{figure}
  \includegraphics[width=\columnwidth,clip,trim=0 1in 0 0]{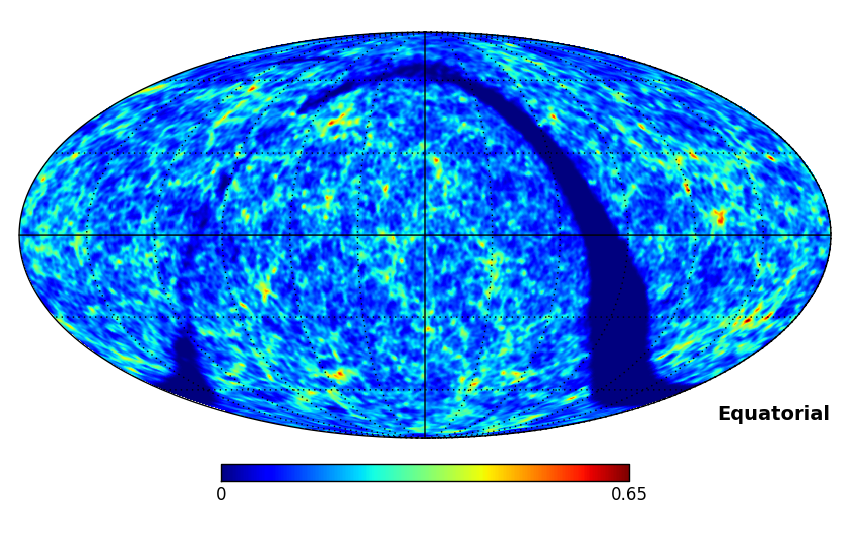}
  \includegraphics[width=\columnwidth,clip,trim=0 1in 0 0]{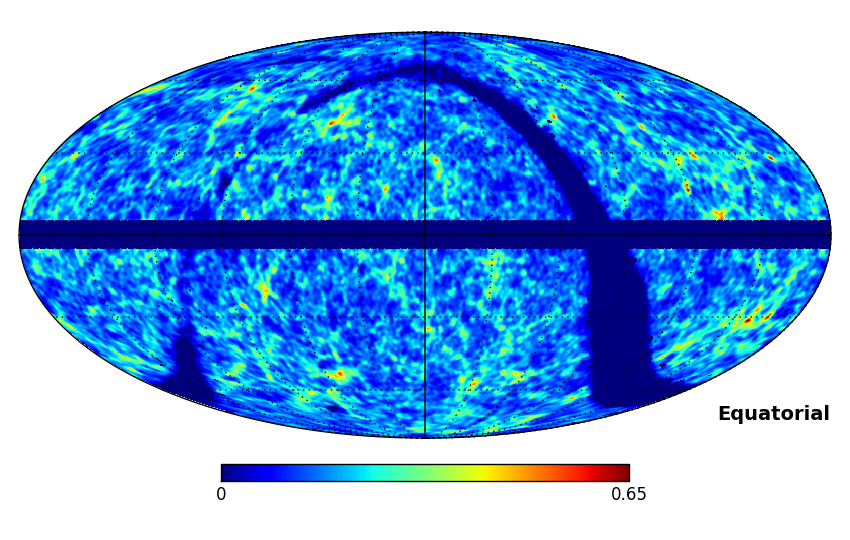}
  \includegraphics[width=\columnwidth,clip,trim=0 1in 0 0]{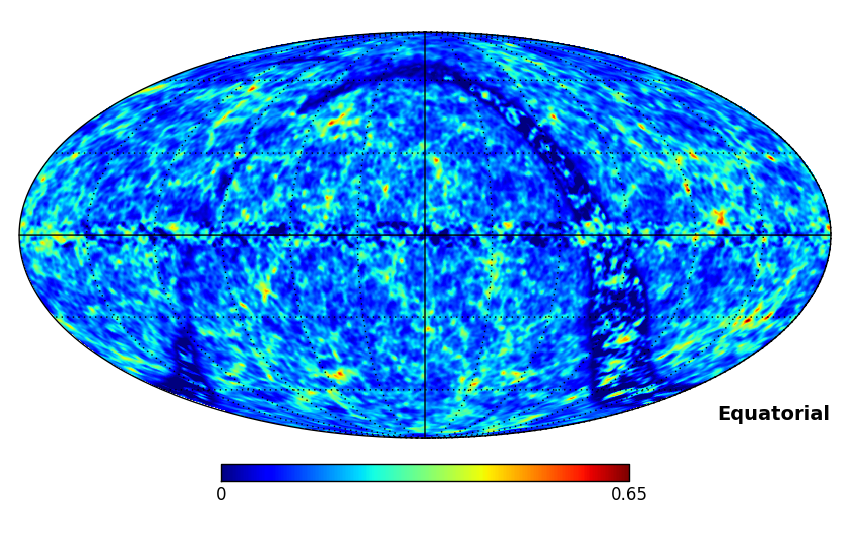}
  \includegraphics[width=\columnwidth,clip,trim=0 0.4in 0 0]{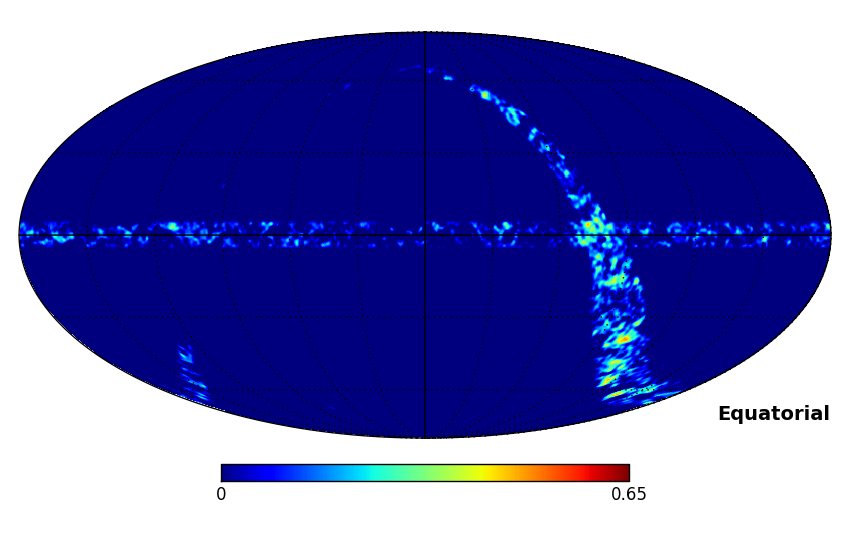}
  \caption{Top: the data used for the infilling test before masking
    (this is the same as Fig.~\ref{fig:galmap} but in equatorial
    coordinated.  Upper middle: we masked both the Galactic plane and within
    five degrees of the celestial equator.  Lower middle: the infilled galaxy
    distribution both in the Galactic plane and the equatorial plane
    to compare with the upper panel. Bottom: the difference between the initial map (top) and the
    infilled map (lower middle).}
  \label{fig:infilling_test}
\end{figure}

To demonstrate its efficacy here, we will first apply the procedure to
a galaxy map that has an additional mask as depicted in
Fig.~\ref{fig:infilling_test}.  We have masked both the Galactic plane
and the equatorial plane.  These equatorial region outside the
Galactic plane is our test region where we know the underlying galaxy
distribution, and we attempt to reconstruct it from the data outside
the masked regions.  Most of the structures within the equatorial
region in the top panel are reproduced in the lower middle panel.  The
difference between the input map and the infill map is depicted in the
bottom panel. To make statistic sense of the agreement we calculate
Pearson's correlation coefficient ($r$) between the original data and
the infilled reproduction within the infilled region outside of the
Galactic plane.  We obtain a value of $r=0.25$.  To estimate the
significance of this value, we perform two tests.  First, we shuffled
the data within the test region and recalculated $r$ for these
shuffled sets.  Over one thousand trials the maximum value of $r$
obtained was 0.0057 and the distribution was consistent with a normal
distribution with $\sigma=0.002$ and a mean of zero or approximately
the reciprocal of the square root of the number of pixels within the
test region.  For an second more stringent test we calculated the
angular power spectrum of the original galaxy map and generated 1,000
galaxy maps consistent with this power spectrum.  This accounts for
the fact that neighbouring regions of sky are correlated, which the
standard shuffle test neglects.  The largest obtained was 0.171, and
the distribution was consistent with a normal distribution with
$\sigma=0.066$ and zero mean, so the observed correlation over the
test region reaches nearly four-sigma significance.  For comparision
the correlation coefficient of the galaxy map with a bootstrapped
realisation of the same map over the test region is typically much
higher $r=0.97$, so clearly much information is lost in the
reconstruction, but the test reveals that the infilling procedure does
give a good first-order guess at the hidden structures.

The region that we have infilled is still apparent in the infilled map,
whereas for the infilled CMB, the infilled region typically is not
apparent.  We believe that this results from two facts.  First, we are
infilling a narrow region, so that we can get as reliable an estimate
of the background galaxy distribution as possible. For the CMB
infilling one just wants a map that is constrained to be the same
outside the infilled region and has the same statistics as the rest of
the map within the region; one does not need a reliable estimate of
the covered sky, so one can mask a larger portion of the image.  The
second reason is that the galaxy density is strictly positive whereas
at the CMB anisotropies are positive and negative.  We have tried some
simple reparametrisations of the galaxy density, but these either
yield the same artifacts (such as subtracting the mean desnity) or
other difficulties (such as using the logarithm of the density).

After demonstrating the efficacy of the infilling procedure, we now
perform the calculation to infill only through the Galactic plane (we did
infill the Galactic plane in the tests as well).  The upper panel of
Fig~\ref{fig:infilling} depicts the mask that we will use to mask the
data in Fig.~\ref{fig:galmap}, and the middle panel gives the initial
galaxy map with the masked region filled in.  There are several
structures within masked region that connect with the structures on either
side of the Galactic plane.  Finally, we can estimate the
signal-to-noise of the infilled map by calculate a series of galaxy
density maps by resampling the 2MPZ to obtain new catalogues, new maps
and new infilled maps.  The lower panel of Fig.~\ref{fig:infilling}
depicts the signal-to-noise ratio of the map.  Outside of the Galactic
plane the signal-to-noise almost everywhere exceeds four.  In the
infilled region most of the overdense structures correspond to high
signal-to-noise regions and therefore may provide a reliable estimate
of the regions in the zone of avoidance where $P(\mathrm{Position})$
is large.

\section{Results}

To assess the performance of these techniques we will first focus on optical
follow-up where we do not wish to observe through the zone of
avoidance.  We will use the Bayesian probability region calculated by
the BAYESTAR algorithm \citep{2015arXiv150803634S} from
\citet{2014ApJ...795..105S} for a LIGO-only detection, that is, before
Virgo is operational.  For simplicity we focus on fields of view that
correspond to a particular valid value of \texttt{NSIDE} for the
HEALPix map.  In particular we examine a 13-square-degree field of
view ($\mathtt{NSIDE}=16$) similar to that of LSST, 0.8
($\mathtt{NSIDE}=64$) and 0.05-square-degree fields of view.  We
quantify the performance in two ways: the time to create the optimised
observing plan is typically 1-3 seconds and the decrease in the number
of fields required to reach a given cumulative probability.

\begin{figure}
  \includegraphics[width=\columnwidth,clip,trim=0 1.35cm 0
    0]{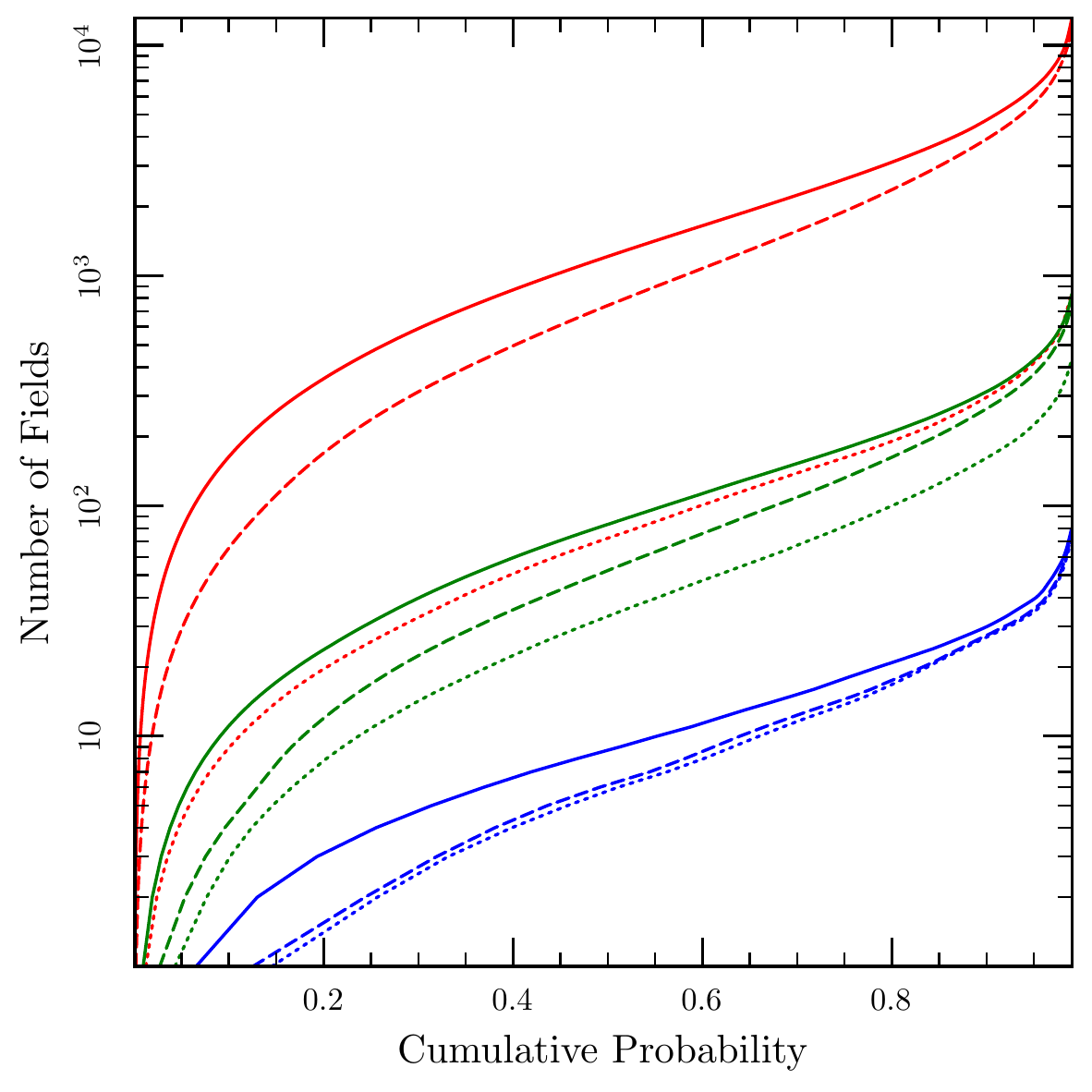}
  \includegraphics[width=\columnwidth,clip,trim=0 0 0
    0.15cm]{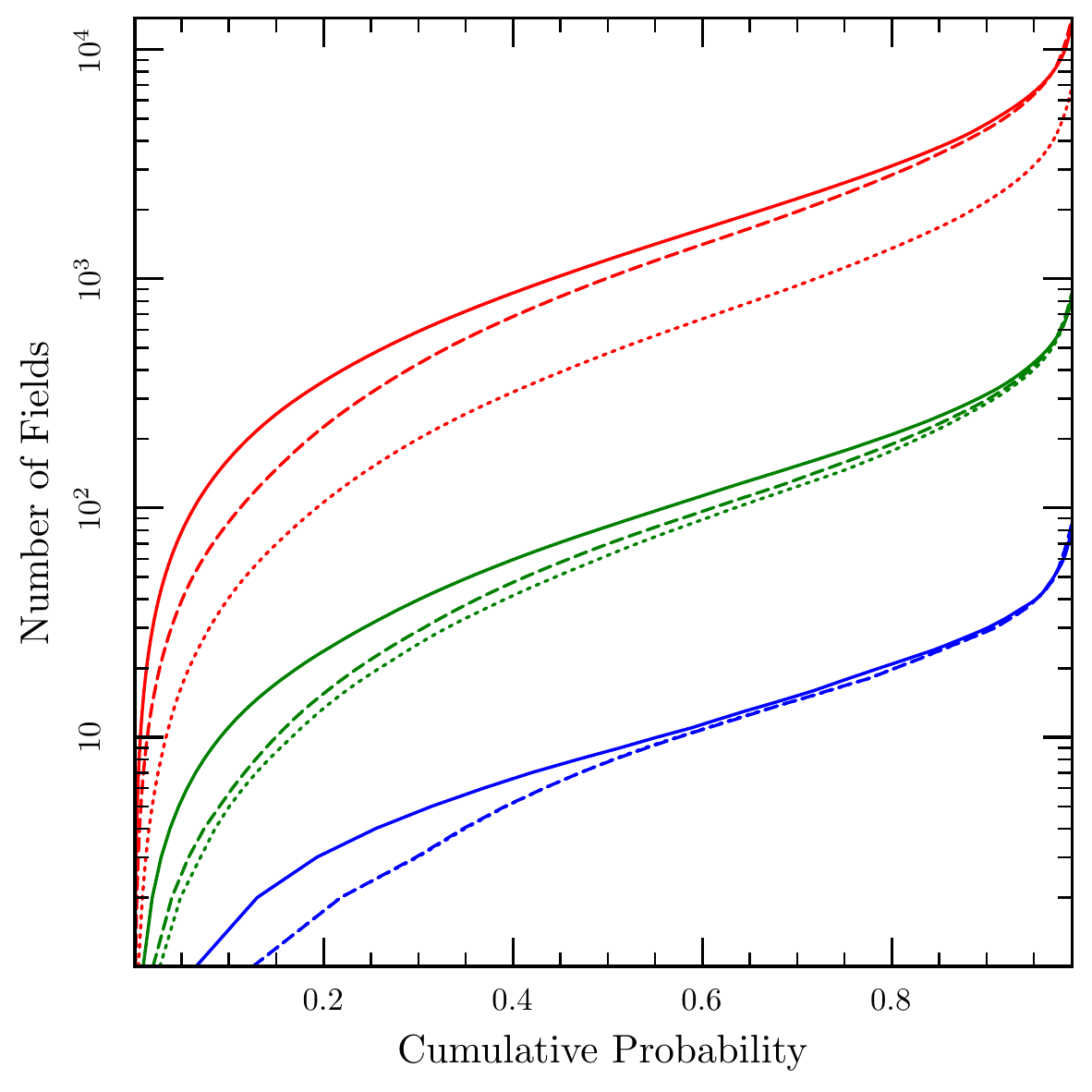}
  \caption{The number of fields required to cover the given fraction
    of the probability region for a simulated LIGO detection (solid curves
    curves without the galaxy map, dashed curves with a smoothed
    galaxy, dotted with a raw galaxy map).  The upper red
    curves use a HEALPix map with about 800,000 cells, the green
    curves have about 50,000 cells, the blue curves have about
    3,000 cells,
    corresponding 0.05, 0.8  and 13 square-degree fields of
    view. The redshift range of the galaxy map in the upper panel is
    $0.03<z<0.04$ and $0.01<z<0.1$ in the lower panel.}
  \label{fig:bayestar}
\end{figure}

Fig.~\ref{fig:bayestar} depicts the results for the different sizes of
fields and the possibility of using an raw (unsmoothed) and smoothed
galaxy map.  The upper panel gives the performance with a galaxy map
restricted to the redshift range $0.03<z<0.04$.  Here the improvement
in the number of fields to observe is most dramatic.  We begin with
the lowest triplet of curves that correspond to the largest field of
view.  Here the improvement is of using a galaxy map is modest, the
number of fields to achieve a given cumulative probability decreases
by about 20\%.  This is because most 13-square-degree regions of the
sky contain a nearby galaxy. Furthermore, with such a large field of
view using a smoothed galaxy map does not affect the results
significantly.  On the other hand, if one uses the alternative metric
of what is the probability of that the source lies within the first
field, the use of a galaxy map increases this probability from about
6.6\% to 14.5\%.

If we now examine the most modest field of view, the 0.05 square-degree
field, we can see a more dramatic advantage of using the galaxy map.
If one uses the raw galaxy map in which each observed field must
contain at least one galaxy, it requires 73 fields (or about 4 square
degrees) to reach half of the cumulative probability.  To reach the
same cumulative probability requires 742 fields (about 38 square
degrees) if one uses the smoothed map and 1,211 fields (about 61 square
degrees) without a galaxy map.  For such a small field of view the
effectiveness of the galaxy map is dramatic.  Furthermore, the chance
of the source being in the first observed field increases from 0.07\%
to 1.2\% with the unsmoothed map.  Understandably for the
intermediate-sized fields of view the improvement is intermediate
between that achieved for the LSST field of view and for the modest one.

If the redshift estimate for the source is much less accurate perhaps
$0.01<z<0.1$, the gains to be had by using a galaxy map are more
modest as depicted in the lower panel of Fig.~\ref{fig:bayestar}.  The
galaxy map corresponding to this range of redshift is given in
Fig.~\ref{fig:galmap}.  For the 13 square-degree field of view, the
improvement is especially modest; with the galaxy map eight fields are
required to reach the fifty percent mark and without the map nine
fields are required.  For the smallest field of view, the galaxy map
reduces the number of fields required to reach the fifty-percent mark
from 1,211 to 473, a 61\% reduction.  With the more accurate redshift
estimate the reduction was nearly 94\%.  The probability of the source
lying in the first field increases from 0.07\% to 0.46\%.  This
stresses the importance of having distance estimates as early as
possible in the data analysis following a burst.

\begin{figure}
  \includegraphics[width=\columnwidth]{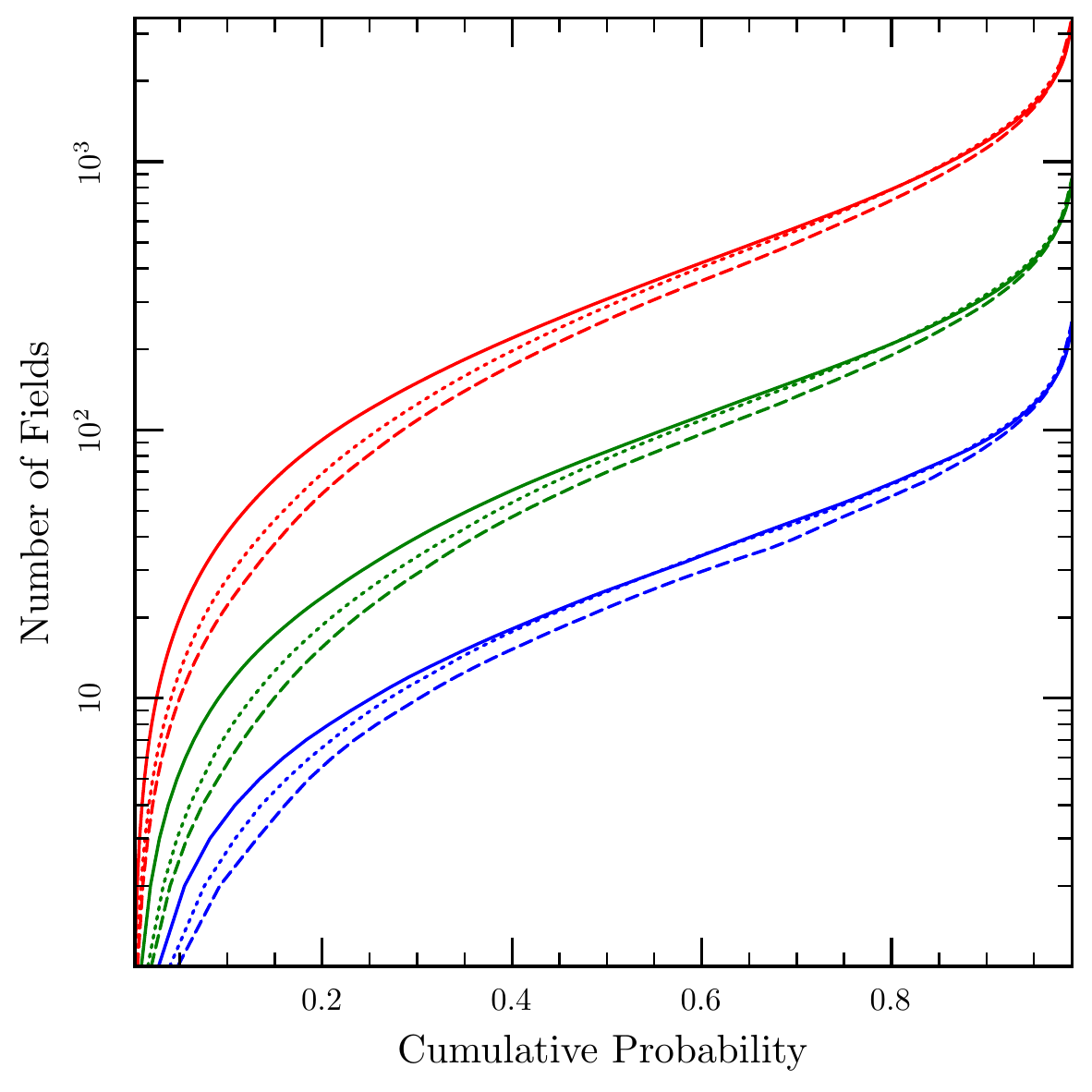}
  \caption{The number of fields required to cover the given fraction
    of the probability region for a simulated LIGO detection: solid
    curves curves without the galaxy map, dashed curves with a
    smoothed galaxy map and the dotted curves with a smoothed maps
    infilled through the Galactic plane.  The upper red curves use a
    HEALPix map with about 200,000 cells, the green curves have about
    50,000 cells, the blue curves have about 12,000 cells,
    corresponding 0.2, 0.8 and 3 square-degree fields of view. The
    redshift range of the galaxy map is 0.01 to 0.1. The green dashed
    and solid curves are the same as those in the lower panel of
    Fig.~\ref{fig:bayestar}.}
  \label{fig:bayestar_inf}
\end{figure}

Fig.~\ref{fig:bayestar_inf} depicts the results using the infilled
galaxy map.  An infilled map is useful to search for electromagnetic
counterparts that may be seen through the Galactic plane, for example,
gamma-rays, hard x-rays and radio emission.  Here we use a narrower
range of fields of view: 0.2, 0.8 and 3 square-degrees.  For example
the Parkes Multibeam receiever \citep{1996PASA...13..243S} can cover
about 3 square-degrees in a single pointing and the XMM-Netwon EPIC
instrument has a 0.25 square-degree field.  Here we will focus on the
middle set of green curves that are for the 0.8-square-degree field of
view, the same as the green curves in lower panel of
Fig.~\ref{fig:bayestar}.  In both figures the upper solid curve gives
the number of fields without a galaxy maps and the dashed curve gives
the number of fields with a smoothed galaxy map.  For these curves,
the first field has one-percent probabilty without a galaxy map and
two percent with the smoothed galaxy map.  The infilled map because it
also has probablity through the Galactic plane yields an intermediate
result of 1.6\% in the first field. To reach half of the probability
requires 84 fields without a galaxy map, 79 with the infilled map and
70 with the smoothed map, so again using the infilled map and
searching through the Galactic plane yields an intermediate result.
One could reduce the number of fields to reach the fiftieth percentile
slightly by using the infilled, smoothed map in the Galactic plane and
the raw galaxy counts outside of the plane.  The total number of
fields required to search half of the probability is 63 for the raw
map, so the improvement achieved through this additional complication
is modest.

\section{Conclusions}

The discovery of gravitational waves from binary-black-hole mergers
highlights the need for rapid three-dimensional localisation of
gravitational-wave events to understand the astrophysical nature of
these sources
\citep{2013ApJ...767..124N,2016arXiv160307333S,2016arXiv160504242S}
Although the Fermi GBM did see hints of gamma rays coincident with the
gravitational wave event in time \citep{2016arXiv160203920C}, other
efforts at finding an electromagnetic counterpart only provided upper
limits
\citep[e.g][]{2016arXiv160208492A,2016arXiv160204198S,2016arXiv160204156S,2016arXiv160204488F},
possibly due to the rapid decay of the electromagnetic radiation from
the binary black-hole merger.  Only the GBM experiment though had
observations coincident in time with the event.  The Fermi LAT managed
to observe the entire probability region within 70 minutes of the
event, so the timescale for rapid follow-up of these events is minutes
rather than hours.  The algorithms presented in this paper offer a
technique to maximise the potential for discovery with a very modest
computation of the fields to observe first and which order.  The key
is to calculate the likely regions of sky to observed before the burst
in the form of a HEALPix map, so at the burst one can rapidly
construct the observing plan.

The information that one puts into the map, of course, will depend on
the nature of the gravitational-wave event ({\em e.g.} its distances,
the masses and composition of the components, {\em etc.}).  Here we
have simply used the surface density of nearby galaxies for the
$P(\mathrm{position})$ map.  However, one can do much better by using
the predicted rates of the various types of events and the types of
galaxies that one expects to find them in.  For example,
\citet{2016arXiv160204531B} argue that an event like GW150914 resulted
from a binary black hole whose progenitor stars formed around a
redshift of 3 (70\% likely) or around a redshift of 0.2 (30\%).  This
information informed by stellar population synthesis could be used to
develop a better guess for the galaxy map by increase the weight of
galaxies whose stars were born during these epochs.
\citet{2015ApJ...806..263D} argue that the coalescence of black-hole
binaries like GW150914 will dominate the detection rates, and
furthermore, most of these binaries form in low-metallicity galaxies
at about 1~Gyr after the Big Bang.  Knowing in which local galaxies
these systems typically end up would greatly improve the follow-up
strategy.  Are we interested in looking at systems that are still low
metallicity with star formation long ago like dwarf spheroidals?  Or
do we expect these systems to have been incorporated in larger
galaxies subsequently and are these larger galaxies typically spirals
or ellipticals today?

The dawn of gravitational wave astronomy is upon us.  To realise its
full potential we can use all of our prior knowledge of the expected
rates of these event in the context of the hierarchical formation of
galaxies to determine where to look for the electromagnetic signal
that hopefully accompanies these events.

The software and galaxy maps used in this paper is available at
\url{http://ubc-astrophysics.github.io}.  We used the VizieR Service,
the NASA ADS service, the SuperCOSMOS Science Archive, the NASA/IPAC
Infrared Science Archive, the HEALPy libraries and arXiv.org. This
work was supported by the Natural Sciences and Engineering Research
Council of Canada, the Canadian Foundation for Innovation, the British
Columbia Knowledge Development Fund and the Bertha and Louis Weinstein
Research Fund at the University of British Columbia.  We would also
like to thank the anonymous referee for support and useful comments.

\bibliography{obsplan}
\bibliographystyle{apj}

\label{lastpage}
\end{document}